\documentclass[aps,preprint,nofootinbib,showpacs]{revtex4}
\usepackage{graphics}
\usepackage{slashed}
\usepackage{amssymb}
\usepackage{lscape}
\usepackage{amsmath}
\usepackage{rotating}
\usepackage{graphicx}
\graphicspath{ {images/} }
\begin{document}

\title{Relativistic quantum geometry from a 5D geometrical vacuum: gravitational waves from preinflation}

\author{$^{1,2}$ Mauricio Bellini, $^{3}$ Jos\'e Edgar Madriz Aguilar \thanks{E-mail address: madriz@mdp.edu.ar},  $^{3}$ M. Montes\thanks{E-mail address: mariana.montnav@gmail.com} and $^{1}$ P. A. S\'anchez\thanks{E-mail address: pabsan@mdp.edu.ar}}
\affiliation{$^{1}$ Departamento de F\'isica, Facultad de Ciencias Exactas y Naturales, Universidad Nacional de Mar del Plata, Funes 3350, C.P. 7600, Mar del Plata, Argentina\\
$^{2}$ Instituto de Investigaciones F\'isicas (IFIMAR), Consejo Nacional de Investigaciones Cient\'{\i}ficas y T\'ecnicas (CONICET), Funes 3350, C.P. 7600, Mar del Plata, Argentina. \\
\\
$^{3}$ Departamento de Matem\'aticas, Centro Universitario de Ciencias Exactas e ingenier\'{i}as (CUCEI),
Universidad de Guadalajara (UdG), Av. Revoluci\'on 1500 S.R. 44430, Guadalajara, Jalisco, M\'exico.  }

\begin{abstract}
In this work we introduce Relativistic Quantum Geometry (RQG) on a Modern Kaluza-Klein theory by studying the boundary conditions on a extended Einstein-Hilbert action for a 5D vacuum defined on a 5D (background) Riemannian manifold. We introduce a connection which describes a displacement from the background manifold to the extended one, on which the 5D vacuum Einstein equations with cosmological constant included, describes the dynamics of the scalar field $\sigma$, which is responsible of describing the mentioned displacement and complies with a relativistic quantum algebra that depends on the relativistic observers. In our formalism the extra dimension is considered as space-like, and therefore is a noncompact one. After considering a static foliation on the extra dimension, we obtain the dynamics for the gravitational waves that propagates on a 4D (curved) background, defined on the 4D induced curved Riemannian manifold. Finally, an example, in which we study a pre-inflationary model of the early universe is developed. We obtain some constraints from Planck2018 observations.
\end{abstract}

\maketitle

\vskip .5cm
 Weyl-Integrable geometry, Riemannian-geometry, quantum geometry, gauge invariance.

\section{Introduction}

The formulation of a theory of quantum gravity free of problems is one of the greatest challenges of modern physics. Nowadays there are many theories attempting to describe gravity at quantum level.  The M string theory \cite{BB1,BB2} and several approaches in the context of loop quantum gravity as spin foam networks, are good examples \cite{BB3,BB4,BB5}. A feature shared by these theories is the use of geometry in different facets. Quantum geometrodynamics, introduced by Wheeler  \cite{BB6,BB7}, and quantum geometry \cite{BB4,BB5} are some of the geometrical frameworks involved. \\

Until now, it has not been possible to formulate a consistent quantum gravity theory that allows the unification of gravity with the other interactions of the standard particle model. An attempt that address this issue is the recently introduced relativistic quantum geometry \cite{M1,M2}. In this approach it is obtained a gauge-invariant  relativistic quantum geometry by using a Weylian-like manifold \cite{M1,M2}. Such manifold is thus endowed with a geometric scalar field which provides a gauge invariant relativistic quantum theory in which the algebra of the Weylian scalar field depends on the observers. In this last framework have been studied several topics. Some of them are inflationary back reaction effects \cite{BB8}, charged and electromagnetic fields \cite{BB9} and geometric back-reaction in pre-inflationary scenarios \cite{BB10}, among others.\\

In this work we extend the formalism of the relativistic quantum geometry to the five-dimensional (5D) scenario of the induced matter theory. At the beginning of the nineties Paul Wesson and collaborators introduced a theory in which our our-dimensional (4D) universe can be locally and isometrically embedded into a 5D space-time in vacuum, where the fith extra dimension is space-like and non-compact \cite{BB11,BB12,BB13}. The letter is organized as follows. In section I we give a little introduction. Section II is dedicated to the formulation of 5D of the relativitic quantum geometry formalism. Section III is devoted to the 5D vacuum case of the formalism developed in section II. In the section IV we show how to obtain an effective 4D setting induced from 5D vacuum. In section V we give as an application of the formalism a model of pre-inflation. In section VI we study the 4D induced gravitational waves during pre-inflation. Finally we leave section VII for some final comments.

\section {The 5D formalism}

 Let us start considering the 5D action in vacuum
 \begin{equation}\label{eq1}
 ^{(5)}\!{\mathcal S}=\frac{1}{2\kappa}\int_{V}d^{5}y\,\sqrt{|g_5|}\,^{(5)}\!R,
 \end{equation}
where $g_5$ is the determinant of the 5D metric $g_{ab}$, $\kappa$ is the 5D gravitational coupling, $^{(5)}\!R$ is the 5D Ricci scalar of curvature and $V$ denotes the volume of the space-time manifold ${\mathcal M}$ which in our case is assumed to has a boundary $\partial{\mathcal M}$. Following a Hilbert variational procedure with respect to the metric tensor,  we arrive to
\begin{equation}\label{eq2}
\delta\,^{(5)}\!{\mathcal S}=\frac{1}{2\kappa}\int d^{5}y\sqrt{|g_5|}\left[\delta g^{ab}\,^{(5)}\!G_{ab}+g^{ab}\delta \,^{(5)}\!R_{ab}\right]=0,
\end{equation}
with $g^{ab}\delta\,^{(5)}\!R_{ab}=\,^{(5)}\!\nabla_c\delta W^{c}$, being $\delta W^{c}=\delta\,^{(5)}\!\Gamma^{c}_{ab}g^{ab}-\delta\,^{(5)}\!\Gamma^{d}_{ed}g^{ec}=g^{ab}\,^{(5)}\!\nabla^{c}\delta \,^{(5)}\!\Psi_{ab}$ where $^{(5)}\!\nabla$ denotes the 5D affine connection. In general, when a bounded manifold is considered, a York-Gibbons-Hawking action is introduced to ensure that the Hilbert variational procedure is well-defined. In our case, we prefer maintaing the boundary term and investigate the implications in the field dynamics due to the presence ot this term. With this idea in mind, when we implement $\delta{\mathcal S}=0$ let us consider the condition: $^{(5)}G_{ab}=\Lambda_5 g_{ab}$, where $\Lambda_{5}$ is denoting the 5D cosmological constant. This condition is in agreement with the idea that our start point in the 5D manifold is a geometrical vacuum. In addition, if we require that $g^{ab}\delta\,^{(5)}\!R_{ab}=\,^{(5)}\!\nabla_c\delta W^{c}=\delta \,^{(5)}\!\Phi$, the equation \eqref{eq2} leaves to the condition $g_{ab}\delta \,^{(5)}\!\Phi=\Lambda_5 \delta g_{ab}$, where $^{(5)}\Phi$ is a 5D scalar field. Then, we postulate the existence of a tensor field $\delta \,^{(5)}\!\Psi_{ab}$ such that $\delta \,^{(5)}R_{ab}\equiv \,^{(5)}\!\nabla_{b}\delta W_{a}-\delta\,^{(5)}\!\Phi g_{ab}\equiv \,^{(5)}\!\Box \delta\,^{(5)}\!\Psi_{ab}-\delta \,^{(5)}\!\Phi g_{ab}=0$. Thus, it follows that
\begin{equation}\label{eq3}
\delta W^{c}=g^{ab}\,^{(5)}\nabla^{c}\delta\,^{(5)}\!\Psi_{ab},
\end{equation}
where $^{(5)}\nabla^{c}\delta\,^{(5)}\!\Psi_{ab}=\delta\,^{(5)}\Gamma^{c}_{ab}-\delta^{c}_{b}\delta\,^{(5)}\!\Gamma^{e}_{ae}$. The fields $\delta W_c$ and $\delta\,^{(5)}\!\Psi_{ab}$ result to be invariant under the gauge transformations
\begin{equation}\label{eq4}
\delta \bar{W}_{a}=\delta W_{a}-\,^{(5)}\!\nabla_{a}\delta\,^{(5)}\!\Phi,\qquad \delta\,^{(5)}\!\bar{\Psi}_{ab}=\delta\,^{(5)}\!\Psi_{ab}-g_{ab}\delta\,^{(5)}\!\Phi,
\end{equation}
where the condition $^{(5)}\!\Box\delta\,^{(5)}\!\Phi=0$ must hold. On the other hand, the field equations and the equation of motion for 5D  gravitational waves can be written as
\begin{eqnarray}\label{eq5}
&& ^{(5)}\bar{G}_{ab}=\,^{(5)}\!G_{ab}-\Lambda_5 g_{ab}=0,\\
\label{eq6}
&& ^{(5)}\!\Box \,\delta\,^{(5)}\!\bar{\Psi}_{ab}=\Box \,\delta\,^{(5)}\!\Psi_{ab}-g_{ab}\delta\,^{(5)}\!\Phi=0,
\end{eqnarray}
where in addition $\delta\,^{(5)}\!\Phi=(\Lambda/5)g^{ab}\delta g_{ab}$ and $^{(5)}\!\Box \equiv \nabla_a\nabla^a$ is the 5D D'Alambertian on the 5D Riemann (background) manifold. The scalar field $\delta\,^{(5)}\Phi(y^{a})$ can be interpreted as a scalar-flux of $\delta W^{a}$ through the 4D boundary $\partial{\mathcal M}$. This flux can be considered as a gravitodynamic potential related to the gauge invariance of $\delta W^{a}$ and $\delta\,^{(5)}\!\Psi_{ab}$. Until now, we have implemented and exact variation of the Einstein-Hilbert action on a Riemannian background geometry. Now the idea is to investigate the possibility to extend the previous formalism for a Weylian-like manifold and associate a relativistic quantum dynamics of the Weylian scalar field by using the fact that $\Lambda_{5}$ is a relativistic invariant. In order to do so we proceed as follows.\\

We consider a Weyl-type geometry  with an alternative covariant derivative where the non-metricity condition is given by: $g_{ab|c}=-\frac{1}{4}\,\left[\sigma_{a}\,g_{cb}+\sigma_{b}\,g_{ac}\right]$, where ``$|a$'' is denoting the new Weyl-like covariant derivate. The Weyl-like affine connection related with this new derivative has the form
\begin{equation}\label{eq7}
^{(5)}\Gamma^{a}_{bc}=\,^{(5)}\!\left\lbrace\,^{a}_{bc}\right\rbrace + \frac{1}{4} \,g_{bc}\sigma^{a},
\end{equation}
where $\sigma^{a}\equiv \sigma^{,a}$ and the first term is denoting the 5D Levi-Civita connection. In the Riemann geometry the identity $\Delta g_{ab}=\,^{(5)}\!\nabla_{c}g_{ab}\,dy^c=0$ is valid. However, in the Weylian-like space-time the variation must be done with \eqref{eq7}. Thus, we obtain
\begin{equation}\label{eq8}
\delta g_{ab}=g_{ab|c}dy^{c}=-\frac{1}{4}\,\left[\,^{(5)}\sigma_b g_{ca}+\,^{(5)}\!\sigma_{a}g_{cb}\right]dy^{c},
\end{equation}
where if we define the operator
\begin{equation}\label{eq9}
\overset{\smile}{y}(t,\bar{r},l)=\frac{1}{(2\pi)^{3/2}}\int d^{3}k_rdk_l \overset{\smile}{e}^{a}\left[b_{k_rk_l}\overset{\smile}{y}_{k_rk_l}(t,\bar{r},l)+b_{k_rk_l}^{\dagger}\overset{\smile}{y}_{k_rk_l}^{*}(t,\bar{r},l)\right],
\end{equation}
with $b_{k_rk_l}^{\dagger}$ and $b_{k_rk_l}$ being the creation and annihilation operators of the 5D space-time manifold, such that $\left<B\left|[b_{k_rk_l},b_{k_rk_l}^{\dagger}]\right|B\right>=\delta^{(3)}(\bar{k}_r-\bar{k}_r^{\prime})\delta (k_l-k_l^{\prime})$ and $\overset{\smile}{e}^{a}=\in ^{a}_{bcdf}\overset{\smile}{e}^{b}\overset{\smile}{e}^{c}\overset{\smile}{e}^{d}\overset{\smile}{e}^{f}$, then we obtain
\begin{equation}\label{eq10}
\left.dy^{a}|B\right>=\left.\,^{(5)}\!U^{a}dS_5|B\right>=\left.\delta\overset{\smile}{y}^{a}(y^b)|B\right>,
\end{equation}
which defines the eigenvalue resulting of applying the operator $d\overset{\smile}{y}^{a}$ on a background quantum state $\left|B\right>$, defined on the Riemannian manifold and where $^{(5)}U^{a}$ is the $5$-velocity in the Riemannian manifold. The 5D line element corresponding to the Weyl-type manifold reads
\begin{equation}\label{eq11}
\left<B\left|d\overset{\smile}{y}_{a}d\overset{\smile}{y}^{a}\right|B\right>=\left(\,^{(5)}\!U_a\,^{(5)}\!U^{a}\right)
dS_5^{2}\left<B|B^{\prime}\right>=dS_{5}^2\delta_{BB^{\prime}}.
\end{equation}
This 5D line element can be interpreted as it provides the displacement of 5D quantum trajectories with respect to the ``classical"
(Riemannian) ones. Thus, the parallel transport of some vector $V^{a}$ defined on the Weyl-type manifold  is given by
\begin{equation}\label{eq12}
\delta V^{a}=\frac{1}{4}\,\sigma^{a}g_{bc}V^{b}dy^c,
\end{equation}
which implies that
\begin{equation}\label{eq13}
\frac{\delta V^{a}}{\delta S_5}=\frac{1}{4}\,\sigma^{a}V^{b}g_{bc}\,^{(5)}\!U^{c},
\end{equation}
where we have regarded that $\Delta V^{a}=0$, i.e. the variation of $V^{a}$ in the 5D Riemannian manifold is null. Due to \eqref{eq13} is it not a difficult task to verify that
\begin{equation}\label{eq14}
\frac{\delta V^{a}}{dS_5}\frac{\delta V_a}{\delta S_5}=-\frac{1}{16}\,\left[\sigma^{a}\,^{(5)}\!U_{a}\right)\left(V_c\,^{(5)}\!U^{c}\right)\left(\sigma^{a}V_{a}\right]\neq 0.
\end{equation}
This equation indicates that the length of a vector $V^{a}$ is not preserved under parallel transporting on the Weylian manifold. The actions in the Weylian and Riemannian manifolds are related through the expression
\begin{equation}\label{eq15}
{\mathcal S}_{5}=\int d^{5}y\,\sqrt{|g_5|}\,\left[\frac{\,^{(5)}\!\hat{R}}{\,\,\,\,2\kappa_5}+ \,^{(5)}\!\hat{{\cal L}}\right]=\int d^{5}y\,\left[\sqrt{|g_5|}\,e^{-\frac{1}{2}\sigma}\right]\,\left\{\left[\frac{\,^{(5)}\!\hat{R}}{\,\,\,\,2\kappa_5} + \,^{(5)}\!\hat{{\cal L}} \right] \,e^{\frac{1}{2}\sigma}\right\}
\end{equation},
where $^{(5)}\!R$ and $\,^{(5)}\!\hat{{\cal L}}$ are respectively, the 5D Ricci scalar and the lagrangian density in the Riemann manifold. Demanding that $\delta {\mathcal S_5}=0$ we arrive to the relation
\begin{equation}\label{eq16}
-\frac{\delta V_5}{V_5}=\frac{\delta\left[\frac{\,^{(5)}\!\hat{R}}{\,\,\,\,2\kappa_5}+ \,^{(5)}\!\hat{{\cal L}}\right]}{\left[\frac{\,^{(5)}\!\hat{R}}{\,\,\,\,2\kappa_5}+ \,^{(5)}\!\hat{{\cal L}}\right]}=\frac{1}{2}\delta\sigma,
\end{equation}
with $V_5=\sqrt{|g_5|}$ is the 5D volume element in the Riemannian manifold and $\delta \sigma=\sigma_a dy^{a}$. The Ricci tensor in the Weyl-type manifold in terms of the Ricci tensor in the Riemann manifold is given by
\begin{equation}\label{eq17}
^{(5)}\!R_{ab}=\,^{(5)}\!\hat{R}_{ab}+\frac{1}{4}\left[\sigma_{a;b}+\frac{1}{2}\sigma_a\sigma_b-g_{ab}\left(\sigma^{c}_{;c}+\frac{1}{2}\sigma_c\sigma^c \right)\right],
\end{equation}
where we have used the expression
\begin{equation}
^{(5)}\!{\delta{R}}_{bc} = \left(\hat{\delta\Gamma}^{a}_{ba} \right)_{| c} - \left(\hat{\delta\Gamma}^{a}_{bc} \right)_{|a}=
\frac{1}{4}\left[\sigma_{a;b}+\frac{1}{2}\sigma_a\sigma_b-g_{ab}\left(\sigma^{c}_{;c}+\frac{1}{2}\sigma_c\sigma^c \right)\right].
\end{equation}
Thus, for the scalar curvature we have
\begin{equation}\label{eq18}
^{(5)}\!{R}=\,^{(5)}\!\hat{R}-\left(\sigma^{a}_{;a}+\frac{1}{2}\sigma_a\sigma^{a}\right).
\end{equation}
Therefore, the Einstein tensor can be written as
\begin{equation}\label{eq19}
^{(5)}\!{G}_{ab}=\,^{(5)}\!\hat{G}_{ab}+\frac{1}{4}\left[\sigma_{a;b}+\frac{1}{2}\sigma_a\sigma_{b}+g_{ab}\left(\sigma^{c}_{;c}
+\frac{1}{2}\sigma_c\sigma^c\right)\right].
\end{equation}
 Now, it follows from the equations \eqref{eq5} and \eqref{eq19} that in the Riemannian manifold $\Lambda_5$ is a constant. However in the Weyl-like manifold we obtain
 \begin{equation}\label{eq20}
 \Lambda_5(\sigma,\sigma_{a})=-\frac{3}{10}\left(\,^{(5)}\!\hat{\Box}\sigma+\frac{1}{2}\sigma_c\sigma^{c} \right),
 \end{equation}
 where we have made use of the fact that $g^{ab} G_{ab }=-5\,\Lambda_5(\sigma,\sigma_{a})$.  Given this dependence on the geometrical scalar field $\sigma$ of $\Lambda_5$, it is possible to consider $\Lambda_5$ as a functional of $\sigma$. Thus, we can define the 5D geometrical quantum action in the Weyl-like manifold
 \begin{equation}\label{eq21}
 ^{(5)}{\mathcal W}=\int d^5y\sqrt{|g_5|}\,\Lambda_5(\sigma,\sigma_a).
 \end{equation}
 The field equations derived from \eqref{eq21} then read
 \begin{equation}\label{eq22}
 \frac{\delta\Lambda_5}{\delta\sigma}-\,^{(5)}\!\hat{\nabla}_{a}\left(\frac{\delta\Lambda_5}{\delta\sigma_a}\right)=0,
 \end{equation}
 where $^{(5)}\!\hat{\nabla}_{a}$ is the Riemannian covariant derivative  and the variations are defined in the Weyl-like manifold. The geometrical canonical momentum is given by $^{(5)}\Pi^{a}\equiv \frac{\delta\Lambda_5}{\delta\sigma_a}=-\frac{3}{10}\sigma^{a}$, whereas the dynamics of $\sigma$ is governed by
 \begin{equation}\label{eq23}
 ^{(5)}\hat{\Box}\sigma=0.
 \end{equation}
 Therefore, it is not difficult to verify  that
 \begin{equation}\label{eq24}
 ^{(5)}\!\hat{\nabla}_{a}\,^{(5)}\!\Pi^{a}=0.
 \end{equation}
 Hence, for the invariant $^{(5)}\!\Pi^2=\,^{(5)}\!\Pi_a\!\!\,^{(5)}\!\Pi^{a}$ we obtain
 \begin{equation}\label{eq25}
 \left[\sigma,^{(5)}\!\Pi^2\right]=\frac{9}{100}\lbrace \sigma_a\left[\sigma,\sigma^{a}\right]+\left[\sigma,\sigma_a\right]\sigma^{a}\rbrace=0,
 \end{equation}
 where we have used that $^{(5)}\hat{U}^{a}\sigma_a=\,^{(5)}\hat{U}_{a}\sigma^{a}$ and
 the relations
 \begin{eqnarray}\label{eq26}
 \left[\sigma(x,l),\sigma_a(x^{\prime},l^{\prime})\right]=i\,^{(5)}\Theta_a\delta^{(4)}(x-x^{\prime})\delta(l-l^{\prime}),\\
 \label{eq27}
 \left[\sigma(x,l),\sigma^{a}(x^{\prime},l^{\prime})\right]=-i\,^{(5)}\Theta^{a}\delta^{(4)}(x-x^{\prime})\delta(l-l^{\prime}),
 \end{eqnarray}
 with $^{(5)}\Theta^{a}=\hbar\,^{(5)}\!\hat{U}^{a}$ and where $^{(5)}\Theta^2=\,^{(5)}\Theta_{a}\,^{(5)}\Theta^{a}=\hbar^2\,^{(5)}\!\hat{U}_a\,^{(5)}\!\hat{U}^{a}=\hbar^2$, is a 5D relativistic invariant. In addition, we define the Hamiltonian operator
 \begin{equation}\label{eq28}
 ^{(5)}{\mathcal H}=\left(\frac{\delta\Lambda_5}{\delta\sigma_a}\right)\sigma_a-\Lambda_5(\sigma,\sigma_a),
 \end{equation}
such that the eigenvalue equation $^{(5)}{\mathcal H}\left|B\right>=\,^{(5)}\! E\left|B\right>$ holds. It is not difficult to verify that $\delta\,^{(5)}\!{\mathcal H}=0$, which means that the total energy $^{(5)}E$ is an invariant in the extended  manifold. In particular, if we consider a 5D vacuum on the 5D Riemann manifold, we must consider that $^{(5)}E=0$.

\section{Five dimensional vacuum}

In order to consider a 5D-vacuum we shall require, at least, that the background manifold be Ricci-flat: $^{(5)}\!\hat{R}_{ab}=0$. Hence, the equation (\ref{eq17}), must
be written as
\begin{equation}\label{eq17+}
^{(5)}\!R_{ab}=\frac{1}{4}\left[\sigma_{a;b}+\frac{1}{2}\sigma_a\sigma_b-g_{ab}\left(\sigma^{c}_{;c}+\frac{1}{2}\sigma_c\sigma^c \right)\right],
\end{equation}
and therefore, the 5D Einstein equation (\ref{eq19}), would be given by the expression
\begin{equation}\label{eq19+}
^{(5)}\!{G}_{ab}=\frac{1}{4}\left[\sigma_{a;b}+\frac{1}{2}\sigma_a\sigma_{b}+g_{ab}\left(\sigma^{c}_{;c}
+\frac{1}{2}\sigma_c\sigma^c\right)\right],
\end{equation}
where the 5D-vacuum on the Riemann manifold requires that [see equation (\ref{eq18})]:
\begin{equation}\label{eq18+}
^{(5)}\!{R}=-\left(\sigma^{a}_{;a}+\frac{1}{2}\sigma_a\sigma^{a}\right).
\end{equation}
Of course, since the scalar field $\sigma$, is a quantum field, only has sense the expectation values of the equations (\ref{eq17+}), (\ref{eq19+}), and (\ref{eq18+}). In this framework, the right side of (\ref{eq19+}) must be interpreted as the stress tensor due to the $\sigma$-field:
\begin{equation}
-\left(8\pi\,G\right)\,^{(5)}\!{T}_{ab}= 2 \frac{\delta{\,^{(5)}\!\cal{L}}}{\delta g^{ab}} - g_{ab}\,\,^{(5)}\!{\cal{L}},
\end{equation}
where the Lagrangian density is related to the Riemann cosmological constant.
\begin{equation}
\,^{(5)}\!{\cal{L}}=\frac{10}{3\kappa}\,\Lambda_5(\sigma,\sigma_{a})= -\frac{1}{\kappa} \left(\sigma^{a}_{;a}+\frac{1}{2}\sigma_a\sigma^{a}\right),
\end{equation}
and $\kappa=8\pi\,G$. Notice that $\,^{(5)}\!{\cal{L}}= \frac{1}{\kappa}\,^{(5)}\!{R}$.

 \section{The 4D formalism}

 In order to obtain the corresponding 4D formalism derived from the previous one, we consider the 5D line element corresponding to
 the five dimensional canonical metric
 \begin{equation}\label{f4d0}
 ds_{5}^2=F^2(l)\left[\hat{h}_{\alpha\beta}(x^{\nu}) \,dx^{\alpha} dx^{\beta} \right] +\epsilon \frac{1}{l_0^2} \left(\frac{dF(l)}{dl}\right)^2\,dl^2,
 \end{equation}
 where $l$ is denoting the fifth extra coordinate, regarded space-like and non-compact, and $\epsilon=\pm 1$ depending on the metric signature.
 In particular, in this work we shall consider the case $\epsilon=-1$, corresponding to a space-like coordinate. Furthermore, we shall consider that $F(l)$ are the Weierstrass functions that obey the following differential equation:
 \begin{equation}
 \left(\frac{dF(l)}{dl}\right)^2 = 4 F(l)^3 -g_2\,F(l)-g_3,
 \end{equation}
 where $g_2$ and $g_3$ are coefficients. Furthermore, $e_1$, $e_2$ and $e_3$ are the roots of the polynomial $4X^3-g_2\,X-g_3=0$, such that $e_1+e_2+e_3=0$. The equation of motion for $\sigma$: (\ref{eq23}), with the metric (\ref{f4d0}), is
 \begin{equation}
 \left[{\rm ln}\left(\sqrt{g}\right)\right]_{,a} \,\sigma^a + g^{ab}\, \sigma_{,ab} =0,
 \end{equation}
 which can be written as
\begin{equation}\label{mot}
\left[{\rm ln}\left(\sqrt{g}\right)\right]_{,\mu} \,\sigma^{\mu} + g^{\mu\nu}\, \sigma_{\mu,\nu}+ \left[{\rm ln}\left(\sqrt{g}\right)\right]_{,4} \,\sigma^{4} + g^{44}\, \sigma_{4,4}=0,
\end{equation}
such that $\mu,\nu$ run from $0$ to $3$. Now, let us assume that the 5D space-time can be foliated by a family of hypersurfaces $\Sigma_0:l=l_0$, such that on every leaf $\Sigma_0$ the 4D induced line element is given by
\begin{equation}\label{f4d1}
ds_4^2=\hat{h}_{\alpha\beta}(x^{\nu})\,dx^{\alpha}dx^{\beta},
\end{equation}
where $\hat{h}_{\alpha\beta}(x^{\mu})$ is the induced 4D metric. With the help of the Gauss-Codazzi-Ricci embedding equations it is well-known that $^{(5)}\!\hat{R}=\,^{(4)}\!\hat{R}-(\hat{K}^{\mu\nu}\hat{K}_{\mu\nu}-\hat{K}^2)=0$, where $\hat{K}_{\mu\nu}$ is the extrinsic curvature tensor of the induced 4D Riemann manifold and $\hat{K}=\hat{h}^{\alpha\beta}\hat{K}_{\alpha\beta}$. Then, it follows from \eqref{eq18} that
\begin{equation}\label{f4d3}
^{(5)}\!{R}=-(\sigma^{\mu}_{\,\,;\mu}+\frac{1}{2}\,\sigma_{\mu}\sigma^{\mu})-(\sigma^{4}_{\,\,;4}+\frac{1}{2}\,\sigma_{4}\sigma^{4}),
\end{equation}
with $\left.^{(4)}\!{R}=-(\sigma^{\mu}_{\,\,;\mu}+\frac{1}{2}\,\sigma_{\mu}\sigma^{\mu})\right|_{l=l_0}$.

\section{Pre-inflation}

Pre-inflation describes a big-bang theory on a complex manifold, in terms of which the universe describes a background semi-Riemannian expansion,
where the effective 4D line element (\ref{f4d1}), is given by
The line element for this case is
\begin{equation}\label{1}
ds_4^2 = \hat{h}_{\mu\nu} d\hat{x}^{\mu} d\hat{x}^{\nu}= e^{2i\hat\theta(t)} d\hat{t}^2 + a^2(t) \hat{\eta}_{ij} d\hat{x}^i d\hat{x}^j,
\end{equation}
with the signature: $(+,-,-,-)$, $a(t)$ is the scale factor and $\sqrt{\hat{h}} = i a^3 e^{i\hat\theta}$. Here $\hat\theta(t)=\frac{\pi}{2} \frac{a_0}{a}$, with $a\geq a_0$, $t$ is a real parameter time and $H_0=\pi/(2 a_0)=1/t_p$, such that $t_p=5.4 \times 10^{-44} \, {\rm sec}$ is the Planckian time. Notice that the metric (\ref{1})
describes a complex manifold such that, at $t=0$ the space-time is Euclidean, but after many Planckian times, when $\hat\theta \rightarrow 0$, it becomes hyperbolic.

The idea of a pre-inflationary expansion of the universe in which the universe begins to expand through a (global) topological phase transition was proposed in\cite{M2}. In this model was studied the birth of the universe using a complex time $\tau(t)=\int e^{i \hat{\theta}(t)}dt$, such that the phase transition from a pre-inflationary to inflationary epoch was examined using a dynamical rotation of the complex time, $\tau(t)$, on the complex plane. After a particular choice of coordinates, one can define a dynamical variable $\theta$:
$\pi/2 \geq \hat{\theta}(t)>0$, such that it describes the dynamics of the system and it is related with the expansion of the universe
\begin{equation}
\hat\theta(t)=\frac{\pi}{2} e^{-H_0 t}.
\end{equation}
With this choice of coordinates, the effective 4D line element (\ref{1}), takes the form
\begin{equation}\label{m}
ds_4^2 =  \left(\frac{\pi a_0}{2}\right)^2 \frac{1}{\hat{\theta}^2} \left[{d\hat{\theta}}^2 - \delta_{ij} d\hat{x}^i d\hat{x}^j\right].
\end{equation}
If we describe an initially Euclidean 4D universe, that thereafter evolves to a globally hyperbolic one, we must take $\hat{\theta} {\rightarrow } 0$, for a $\hat{\theta}$ that have an initial value $\hat\theta_0=\frac{\pi}{2}$.

\begin{figure}[h]
\noindent
\includegraphics[width=.6\textwidth]{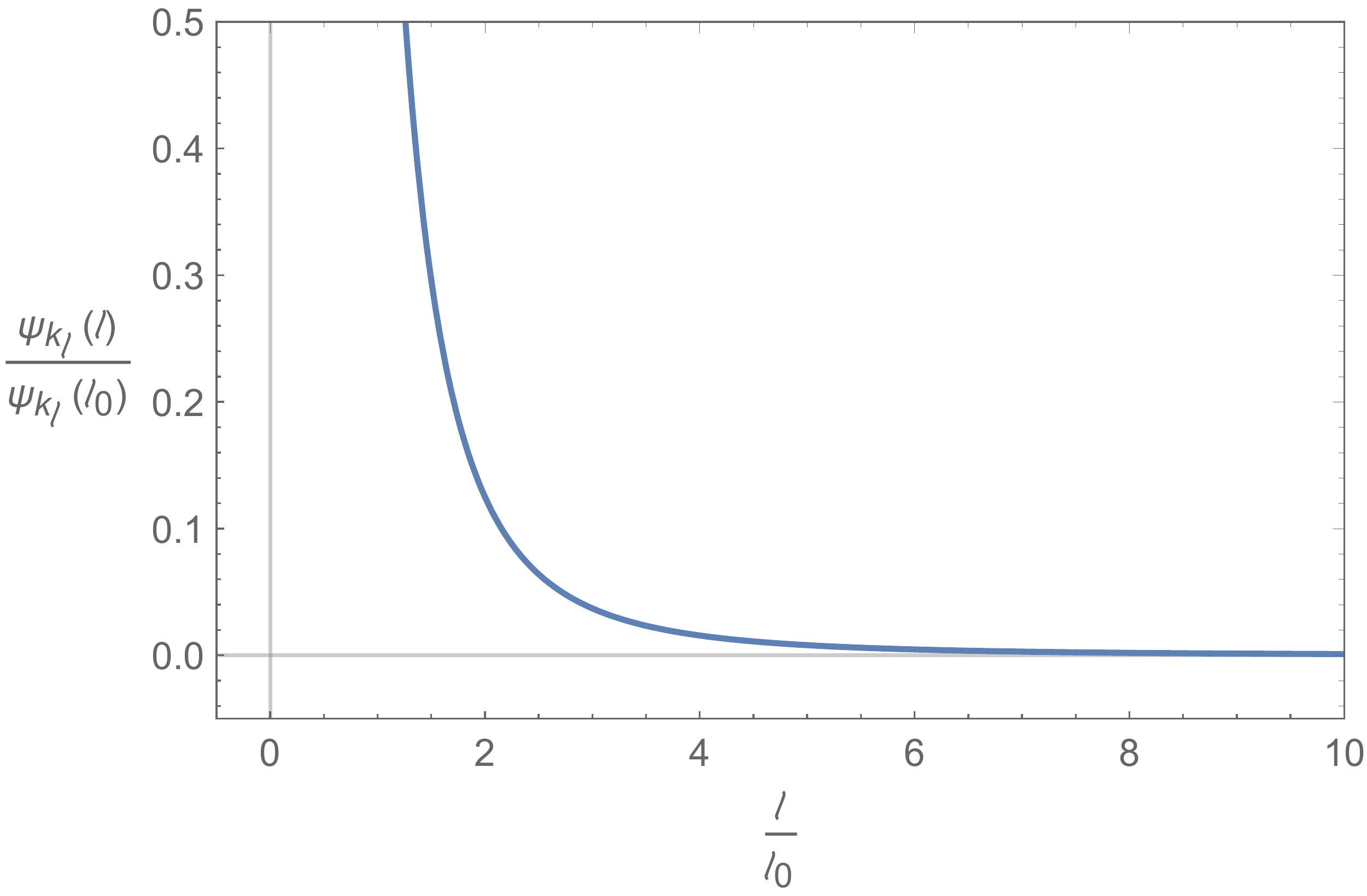}\vskip -0cm\caption{Plot of $\Psi_{k_l}(l)$ for $A=1$, $B=0$, $n=1$, $m\,l_0=10^{-5}$ and $H_0=1/l_0$. Notice that the solution has the limit: ${\Psi_{k_l}(l)}_{l\rightarrow \infty}\rightarrow 0$.}\label{f1}
\end{figure}

The field $\sigma$ can be expanded in terms of a Fourier expansion on the 5D canonical metric (\ref{f4d0})
\begin{equation}
\sigma\left(\theta,\vec{x},l\right) = \frac{1}{(2\pi)^{2}}\,\int d^3k \int dk_l \left[A_{k,k_l}\,e^{i\,\vec{k}.\vec{x}}\,
\Psi_{k_l}(l)\, \xi_k(\theta) + A^{\dagger}_{k,k_l}\,e^{-i\,\vec{k}.\vec{x}} \,\Psi^*_{k_l}(l) \xi^*_k(\theta)\right],
\end{equation}
where $k_l$ is the component of the wavenumber related to the extra dimension. We can propose the separation of variables in order to resolve the equation (\ref{mot}): $\sigma_{k,k_l}\left(\theta,\vec{x},l\right) \sim e^{i\,\vec{k}.\vec{x}}\,
\Psi_{k_l}(l)\, \xi_k(\theta)$. We obtain the following equations for $\Psi_{k_l}(l)$ and $\xi_k(\theta)$:
\begin{eqnarray}
\frac{\partial^2 }{\partial l^2} \Psi_{k_l}(l) & + & \left[\frac{4}{l^4_0}\frac{1}{(F')^3 F} - \frac{F''}{F'}\right] \frac{\partial}{\partial l} \Psi_{k_l}(l) -\left[\frac{m^2}{l^2 (F')^2 F^2}\right] \Psi_{k_l}(l)=0 ,  \label{l} \\
\ddot{\xi}_k &-&   \frac{2}{\hat\theta} \dot{\xi}_k + K^2\, \xi_k(\hat\theta) =0, \label{mm}
\end{eqnarray}
where $K^2=k^2-m^2$, and the {\em dot} denotes the derivative with respect to $\hat\theta$. Notice that $m$ is the induced mass of $\sigma$.
We shall consider the case where
\begin{equation}\label{ff}
F(l)=\left(\frac{l}{l_0}\right)^n,
\end{equation}
with $n>1$. In this case the general solution of (\ref{l}) is
\begin{eqnarray}
\Psi_{k_l}(l) & = & e^{\left[\frac{\left(l/l_0\right)^{-4n} l^4 - l^4_0 \left(l/l_0\right)^{4(1-n)}}{2\,l_0^4 n^3(n-1)}\right]} \left\{ A\, {\cal H}_{yp} \left\{ \left[a_1\right],\left[b_1\right],  \frac{\left(l/l_0\right)^{4(1-n)}}{n^3(n-1)}\right\} \right. \nonumber \\
& + & \left.B\,\left(\frac{l}{l_0}\right)^n\,{\cal H}_{yp} \left\{ \left[a_2\right],\left[b_2\right],  \frac{\left(l/l_0\right)^{4(1-n)}}{n^3(n-1)}\right\} \right\},
\end{eqnarray}
where ${\cal H}_{yp}\left\{ \left[a\right],\left[b\right],  \xi(l)\right\}$ is the hypergeometric function with parameters $a$ and $b$ and argument
$\xi(l)=\frac{\left(l/l_0\right)^{4(1-n)}}{n^3(n-1)}$. In our case the parameters take the form
\begin{eqnarray}
a_1 & = & \frac{n\,m^2\,l_0}{16(n-1)}, \qquad a_2 = \frac{n\,\left[m^2-(4/l^2_0)\right]\,l_0}{16(n-1)}, \nonumber \\
b_2 & = & \frac{5n-4}{4(n-1)}, \qquad b_2 = \frac{3n-4}{4(n-1)}.
\end{eqnarray}
In the figure (\ref{f1}) we show a plot of $\Psi_{k_l}(l)$, where we have taken $n=1$ and the values of the constants $A=1$ and $B=0$. Furthermore the mass of the $\sigma$ is $m\,l_0=10^{-5}$. Notice that the solution is confined on the extra dimension $l$, but the solution tends to $0$ as $l\rightarrow \infty$. \\
The case $n=1$ is more simple and interesting. A particular solution is
\begin{equation}
\Psi_{k_l}(l)=\Psi_{k_l}(l_0)\,\left[\frac{l}{l_0}\right]^{-3/2-1/2\,\sqrt {4\,{m}^{2}{{\it
l_0}}^{2}+9}},
\end{equation}
which tends to $0$ as $\left[\frac{l}{l_0}\right]\rightarrow \infty$.

\subsection{Pre-inflationary 4D background dynamics and  back-reaction effects}

During pre-inflation the universe emerges describing a (global) topological phase transition such that the equation of state is a vacuum expansion. In this case the asymptotic scale factor, Hubble parameter and the
potential are are respectively given by
\begin{equation}
a(t)= a_0\, e^{H_0 t}, \qquad \frac{\dot{a}}{a} = H_0 \qquad V= \frac{3}{8\pi G} H^2_0,
\end{equation}
so that, due to the fact that $\frac{\delta V}{\delta\phi}=0$, the dynamics of $\phi$, is given by the equation of motion
\begin{equation}
\phi{''}-\frac{2}{\hat\theta} \phi{'}=0.
\end{equation}
The solution that describes a field that drives a phase transition of the global geometry
from a 4D Euclidean space to a 4D hyperbolic spacetime, is
\begin{equation}
\phi(t)= \phi_0.
\end{equation}
Furhtermore, the effective 4D energy density and the pressure, are respectively given by
\begin{equation}
\rho(\hat\theta) = \frac{1}{\pi G} \frac{3}{(\pi a_0)^2}, \qquad\qquad
P(\hat\theta) = - \frac{1}{\pi G} \frac{3}{(\pi a_0)^2},
\end{equation}
and therefore, the equation of state for the metric (\ref{m}), is
\begin{equation}
\frac{P}{\rho} =  - 1,
\end{equation}
which describes an effective 4D vacuum expansion during the global topological phase transition (from a 4D Euclidean space to a 4D hyperbolic spacetime) described by $\hat\theta$. The fluctuations of energy density due to the back-reaction effects are
\begin{equation}\label{de}
\frac{1}{\hat{\rho}} \frac{\delta \hat{\rho}}{\delta S} = - 2 \left(\frac{\pi}{2a_0}\right) \, \hat\theta \dot{\sigma},
\end{equation}
such that \cite{M2}, for $\dot{\sigma} = \left< (\dot\sigma)^2 \right>^{1/2}$
\begin{equation}\label{fl}
\left< (\dot\sigma)^2 \right> = \frac{1}{(2\pi)^{3}} \, \int d^3k (\dot{\xi}_k) \, \dot{({\xi}^*_k)},
\end{equation}
where the modes $\xi_k$ must be restricted to
\begin{equation}\label{rcond}
\dot{({\xi}^*_k)} \xi_k - \dot{({\xi}_k)} \xi^*_k= i \hat{\theta}^2 \left(\frac{2}{\pi a_0}\right)^2,
\end{equation}
in order to the field $\sigma$ to be quantized\cite{M1}
\begin{equation}
\left[\sigma(x), \sigma_{\mu}(y)\right] =i \, \hbar \Theta_{\mu} \delta^{(4)}(x-y).
\end{equation}
Here, $\Theta_{\mu}=\left[\hat{\theta}^2 \left(\frac{2}{\pi a_0}\right)^2,0,0,0\right]$ are the components of the background relativistic tetra-vector on the Riemann manifold. The equation of motion for the modes of $\sigma$: $\xi_k(\hat\theta)$, is
The quantized solution of (\ref{mm}) results to be
\begin{equation}
\xi_k(\theta)= \frac{i}{2}\left(\frac{\pi}{2 a_0}\right) K^{-3/2} \,e^{-i K \hat\theta} \left[K\hat\theta-i\right].
\end{equation}
Therefore, the fluctuations (\ref{fl}), are
\begin{equation}
\left< (\dot\sigma)^2 \right> = \frac{1}{8} \frac{\hat\theta^2}{(4 a_0)^2} \epsilon^4 K^4_0,
\end{equation}
such that $\epsilon \ll 1$ and $K_0=\frac{\sqrt{2}}{\hat\theta}$. Hence, the amplitude of energy-density fluctuations on super Hubble scales, becomes
\begin{equation}
\left|\frac{1}{\hat{\rho}} \frac{\delta \hat{\rho}}{\delta S}\right| =  \frac{\pi \epsilon^2}{4\sqrt{2} a^2_0} ,
\end{equation}
which is a constant.

\section{4D gravitational waves during pre-inflation}

In order to obtain 4D  gravitational waves from the 5D Weylian manifold it follows from \eqref{eq6} that
\begin{equation}\label{gwe1}
^{(5)}\Box\,\delta\!\,^{(5)}\Psi_{ab}=0,
\end{equation}
where $^{(5)}\Box\,\delta\!\,^{(5)}\Psi_{ab}=g^{cd}(\delta\!\,^{(5)}\Psi_{ab})_{;dc}$. The equation \eqref{gwe1} is the 5D gravitational wave equation on the 5D Ricci flat Riemann manifold. Proposing the separation of variables $\delta\!\,^{(5)}\Psi_{\alpha\beta}=\Omega(l)\delta\!\,^{(4)}\Psi_{\alpha\beta}(x^{\mu})$ and employing \eqref{f4d0} and  \eqref{m} the equation \eqref{gwe1} for $\delta\!\,^{(5)}\Psi_{\alpha\beta}$ results in the system
\begin{eqnarray}\label{gwe7}
\delta^{(4)}\!\ddot{\Psi}_{\alpha\beta}-\frac{2}{\theta}\delta^{(4)}\!\dot{\Psi}_{\alpha\beta}-\nabla^2\delta^{(4)}\!\Psi_{\alpha\beta}+\left(\frac{\beta^2}{\theta^2}-\alpha\right)\delta^{(4)}\!\Psi_{\alpha\beta}=0,\\
\label{gwe8}
\frac{d^2\Omega}{dl^2}+\frac{n}{l}\frac{d\Omega}{dl}-\frac{\alpha}{l^2}\left(\frac{2nl_0}{\pi a_0}\right)^2\Omega=0,
\end{eqnarray}
where $\alpha$ is a separation constant and $\beta^2=\left(\frac{\pi a_0}{2l_0}\right)^2-\frac{2(n-1)}{n}\left(\frac{\pi a_0}{2l_0}\right)^2-4$. The general solution for \eqref{gwe8} is given by
\begin{equation}\label{gwe9}
\Omega(l)=B_1 l^{-\frac{1}{2}n+\frac{1}{2}+\frac{1}{2}\sqrt{n^2-2n+1+4\alpha\left(\frac{2nl_0}{\pi a_0}\right)^2}}+B_2l^{-\frac{1}{2}n+\frac{1}{2}-\frac{1}{2}\sqrt{n^2-2n+1+4\alpha\left(\frac{2nl_0}{\pi a_0}\right)^2}}.
\end{equation}
 Now, expanding $\delta\!\,^{(4)}\Psi_{\alpha\beta}$ in Fourier modes we obtain
\begin{equation}\label{gwe10}
\delta\!\,^{(4)}\Psi_{\alpha\beta}(\theta,\bar{x})=\frac{1}{(2\pi)^{3/2}}\sum_{M=+,\times}\int d^{3}k\epsilon^{M}_{\alpha\beta}(\hat{z})\left[a_k e^{i\bar{k}\cdot\bar{x}}Q_{k}(\theta)+a_k^{\dagger} e^{i\bar{k}\cdot\bar{x}}Q_{k}^{*}(\theta)\right],
\end{equation}
where $\epsilon^{M}_{\alpha\beta}$ is the polarization tensor and $M=+,\times$ denotes the transverse polarizations, such that $\epsilon^{M}_{\alpha\beta}\epsilon^{\alpha\beta}_{M^{\prime}}=\delta^{M}_{M^{\prime}}$. The annihilation and creation operators $a_k$ and $a_k^{\dagger}$ satisfy the usual commutation algebra
\begin{equation}\label{gwe11}
\left[a_k,a_{k^{\prime}}\right]=\delta^{(3)}(\bar{k}-\bar{k}^{\prime}),\quad \left[a_{k},a_{k^{\prime}}\right]=\left[a_{k}^{\dagger},a_{k^{\prime}}^{\dagger}\right]=0.
\end{equation}
In our formalism we consider tensor perturbations of the metric propagating along the $\hat{z}$-direction, and thus the polarizations are
\begin{equation}\label{gwe12}
\epsilon^{+}_{\alpha\beta}=\left(
\begin{array}{c c}
1 & 0 \\
0 & -1
\end{array}
\right)_{\alpha\beta}\quad
\epsilon^{\times}_{\alpha\beta}=\left(
\begin{array}{c c}
0 & 1\\
1 & 0
\end{array}
\right)_{\alpha\beta},
\end{equation}
with $\alpha\beta$ spanning the $(x,y)$ plane. In order to obtain information about 4D gravitational waves we choose a TT-gauge. Hence, the conditions
\begin{equation}\label{gwe13}
\delta\,^{(4)}\!\Psi_{0\mu}=0,\quad \delta\,^{(4)}\!\Psi^{i}_{i}=0,\quad \hat{\nabla}^{j}\delta\,^{(4)}\!\Psi_{ij}=0,
\end{equation}
are valid.  Therefore, the equation for the $Q_k(\theta)$ modes is given by
\begin{equation}\label{gwe14}
\ddot{Q}_k-\frac{2}{\theta}\dot{Q}_{k}+\left(k^2+\frac{\beta^2}{\theta^2}-\alpha\right)Q_k=0.
\end{equation}
With the help of \eqref{rcond} the normalized solution of \eqref{gwe14} reads
\begin{equation}\label{gwe15}
Q_k(\theta)=\frac{i}{\sqrt{\pi} a_0}\theta^{3/2}{\cal H}_{\lambda}^{(1)}(\sqrt{k^2-\alpha}\theta),
\end{equation}
where ${\cal H}_{\lambda}^{(1)}$ is the first kind Hankel function and $\lambda=\frac{1}{2}\sqrt{9-4\beta}$. Using \eqref{gwe15} the mean squared fluctuations for 4D gravitational waves is then given by
\begin{equation}\label{gwe16}
\left<\delta\,^{(4)}\Psi^2\right>=\frac{2^{2\lambda-1}\Gamma^2(\lambda)}{\pi^5 a_0^2}\theta^{3-2\lambda}\int \frac{dk}{k}k^{3-2\lambda}.
\end{equation}
Most recent observations from Planck2018 constrain the spectral index $n_s=1-n_T=0.965$\cite{planck}, where $n_T=3-2\lambda=0.036$ is the tensor index of the spectrum for gravitational waves. Hence, we obtain
\begin{equation}
n^2_T - 6\,n_T + (\pi\,a_0\,\alpha)^2=0,
\end{equation}
which implies that
\begin{equation}
\alpha = \frac{0.1454}{a_0}=0.092\,H_0,
\end{equation}
where the asymptotic Hubble parameter $H_0$, is related to $a_0$ by the expression: $H_0={\pi\over (2\,a_0)}$. If we take $H_0=1/l_0$, we obtain for the case $n=1$, that
\begin{equation}
\beta^2 = \left(\frac{\pi}{2}\right)^2 -4 \simeq -1.53.
\end{equation}

\section{4D Scalar fluctuations during Pre-inflation}

The equation that describe the scalar fluctuations of the metric is given by
\begin{equation}\label{sflu1}
^{(5)}\Box\delta\,^{(5)}\sigma=0.
\end{equation}
With the help of the line element \eqref{f4d0}, the equation  \eqref{sflu1} acquires the form
\begin{equation}\label{gwe6}
\frac{1}{\sqrt{-h}}\frac{\partial}{\partial x^{\mu}}\left[\sqrt{-h}F^{-2}h^{\mu\nu}\left(\delta\!\,^{(5)}\sigma\right)_{,\nu}\right]-\frac{l_0^2}{F^4\frac{dF}{dl}}\frac{\partial}{\partial l}\left[F^4\left(\frac{dF}{dl}\right)^{-1}\left(\delta\!\,^{(5)}\sigma\right)_{,l}\right]=0.
\end{equation}
Proposing the separation of variables $\delta\!\,^{(5)}\sigma=L(l)\delta\!\,^{(4)}\sigma(x^{\mu})$ and employing \eqref{m} the equation \eqref{gwe6} results in the system
\begin{eqnarray}\label{gwe7}
\delta\!\,^{(4)}\ddot{\sigma}-\frac{2}{\theta}\delta\!\,^{(4)}\dot{\sigma}-\hat{\nabla}^2\delta\!\,^{(4)}\sigma+\left(\frac{\pi a_0}{2}\right)^2\frac{\gamma^2}{\theta^2}\delta\!\,^{(4)}\sigma=0,\\
\label{gwe8}
\frac{d^2L}{dl^2}+\frac{1+3n}{l}\frac{dL}{dl}+\frac{\gamma^2}{(1+3n)l_0^2}\frac{1}{l^2}L=0,
\end{eqnarray}
where $\gamma$ is a separation constant with $length$ units. The general solution for \eqref{gwe8} is given by
\begin{equation}\label{gwe9}
L(l)=E_1 l^{-\frac{3}{2}n}{\bold J}_{\mu}\left[\frac{2\gamma\sqrt{l}}{l_0\sqrt{1+3n}}\right]+E_2 l^{-\frac{3}{2}n}{\bold Y}_{\mu}\left[\frac{2\gamma\sqrt{l}}{l_0\sqrt{1+3n}}\right],
\end{equation}
where $\mu=3n$, ${\bold J}_{\mu}$, ${\bold Y}_{\mu}$ are denoting the Bessel functions and $E_1$, $E_2$ are integration constants. Now, expanding $\delta\!\,^{(4)}\sigma$ in Fourier modes we obtain
\begin{equation}\label{gwe10}
\delta\!\,^{(4)}\sigma(\theta,\bar{x})=\frac{1}{(2\pi)^{3/2}}\int d^{3}k\left[b_k e^{i\bar{k}\cdot\bar{x}}\Theta_{k}(\theta)+b_k^{\dagger} e^{i\bar{k}\cdot\bar{x}}\Theta_{k}^{*}(\theta)\right],
\end{equation}
where the annihilation and creation operators $b_k$ and $b_k^{\dagger}$ satisfy respectively the commutation algebra
\begin{equation}\label{gwe11}
\left[b_k,b_{k^{\prime}}\right]=\delta^{(3)}(\bar{k}-\bar{k}^{\prime}),\quad \left[b_{k},b_{k^{\prime}}\right]=\left[b_{k}^{\dagger},b_{k^{\prime}}^{\dagger}\right]=0.
\end{equation}
Therefore, the equation for the $\Theta_k(\theta)$ modes is given by
\begin{equation}\label{gwe14}
\ddot{\Theta}_k-\frac{2}{\theta}\dot{\Theta}_{k}+\left[k^2+\left(\frac{\pi a_0}{2}\right)^2\frac{\gamma^2}{\theta^2}\right]\Theta_k=0.
\end{equation}
With the help of \eqref{rcond} the normalized solution of \eqref{gwe14} reads
\begin{equation}\label{gwe15}
\Theta_k(\theta)=\frac{i}{\sqrt{\pi} a_0}\theta^{3/2}{\cal H}_{\eta}^{(1)}(k\theta),
\end{equation}
where ${\cal H}_{\eta}^{(1)}$ is the first kind Hankel function and $\eta=\frac{1}{2}\sqrt{9-\pi^2a_0^2\gamma^2}$. Using \eqref{gwe15} the mean squared fluctuations for 4D scalar fluctuations is then given by
\begin{equation}\label{gwe16}
\left<(\delta\,^{(4)}\sigma)^2\right>=\frac{2^{2\eta-1}\Gamma^2(\eta)}{\pi^5 a_0^2}\theta^{3-2\lambda}\int \frac{dk}{k}k^{3-2\eta}.
\end{equation}
It is known that the spectral index is given by $n_s=3-2\eta$, so that we obtain the expression 
\begin{equation}
\left(\gamma a_0\right)^2 = \frac{n_s ( 6-n_s)}{\pi^2},
\end{equation}
which can be estimated by using the recent results for the spectral index\cite{planck}, $n_s\simeq 0.964$:
\begin{equation}
\left(\gamma a_0\right)^2 \simeq 0.492.
\end{equation}
This result means that the constant of separation, $\gamma$ is of the order of $\gamma \simeq {7\over 10\,a_0}$. Due to the fact, $H_0={\pi\over (2\,a_0)}$, we obtain that this constant of separation can be estimated in terms of the Hubble parameter during pre-inflation: $ \gamma \simeq 0.445\, H_0$, which is an important result obtained from the power-spectrum evidence.

\section{Final Comments}

We have studied gravitational waves using a non-perturbative formalism in the framework of the Modern Kaluza-Klein theory, also called Space-Time-Matter theory or Induced Matter theory of gravity, in which the extra dimension is considered as space-like. In this theory it is defined a Ricci-flat background manifold on which we define the back-reaction fluctuations of spacetime. When we take a static foliation on the extra coordinate, we obtain that the effective scalar field $\sigma(x^{\alpha},l=l_0)$, evaluated on the effective 4D space-time, describes a dynamics corresponding to a massive scalar field, with mass $m$. This mass is interpreted as a drag effect on the spacetime $ds_4^2=\hat{h}_{\alpha\beta}(x^{\nu})\,dx^{\alpha}dx^{\beta}$, due to the fact the solution of $\sigma$ on the extra coordinate: $\Psi_{k_l}(l=l_0)$, induces a constant $m$ when we apply separation of variables to resolve the equations. In the figure (\ref{f1}) we have
taken $m\,l_0\equiv m/H_0=10^{-5}$. This is the most relevant difference with respect to the original 4D formalism \cite{M1}. Finally, we have obtained a spectrum of gravitational waves during pre-inflation which fits in Planck 2018 observational data for $\beta^2\simeq -1.53$.

\section*{Acknowledgements}

\noindent M. Bellini and P. A. S\'anchez acknowledge CONICET, Argentina (PIP 11220150100072CO) and UNMdP (EXA852/18) for financial support. J.E.Madriz-Aguilar and M. Montes  acknowledge CONACYT M\'exico, Centro Universitario de Ciencias Exactas e Ingenier\'{\i}as of Universidad de Guadalajara for financial support.

\newpage

\end{document}